\documentclass[a4paper,11pt]{article}
\usepackage{pos}

\title{Constructing {\tt PineAPPL} grids on hardware accelerators}

\author*[a]{Stefano Carrazza}
\author[a]{Juan M. Cruz-Martinez}
\author[a]{Christopher Schwan}

\affiliation[a]{TIF Lab, Dipartimento di Fisica, Universit\`a degli Studi di Milano and INFN, Milan, Italy.}

\emailAdd{stefano.carrazza@unimi.it, juan.cruz@mi.infn.it, christopher.schwan@mi.infn.it}

\abstract{In this proceedings we demonstrate how to implement and construct the
 {\tt PineAPPL} grids, designed for fast-interpolation of Monte Carlo simulation
 with electroweak and QCD corrections, using the {\tt VegasFlow} framework for
 Monte Carlo simulation on hardware accelerators. We provide an example of
 synchronous and asynchronous filling operations of {\tt PineAPPL} grids from
 Monte Carlo events generated by {\tt VegasFlow}. We compare the performance of
 this procedure on multithreading CPU and GPU.}

\FullConference{%
  The Eighth Annual Conference on Large Hadron Collider Physics-LHCP2020 \\
  25-30 May, 2020\\
  online}

\begin{document}
\maketitle

\section{Introduction and implementation}

The fast evaluation of theoretical predictions for a generic set of parton
distribution functions~\cite{Forte:2020yip} and scale variation choices is a
common request that has been addressed by generic tools such as {\tt
APPLGRID}~\cite{Carli:2010rw}, {\tt FastNLO}~\cite{Kluge:2006xs} and more
recently {\tt PineAPPL}~\cite{Carrazza:2020gss,christopher_schwan_2020_3992765}.
In particular, the technology developed by the {\tt PineAPPL} library provides
the possibility to produce fast-interpolation grids of physical cross sections,
computed with a general-purpose Monte Carlo generator, accurate to fixed order
in the strong, electroweak, and combined strong-electroweak couplings.

In order to construct and use {\tt PineAPPL} grids we need to interface the code
to a Monte Carlo simulation library that generates event weights and kinematics
configurations for the grid filling procedure. In this proceedings we use {\tt
VegasFlow}~\cite{Carrazza:2020rdn,juan_cruz_martinez_2020_3691927}, a Monte
Carlo simulation framework with support for multithreading CPU, single-GPU and
multi-GPU setups. The choice of {\tt VegasFlow} relies on high efficiency when
performing simulation thanks to its flexibility to distribute event generation
across multiple hardware accelerators. Furthermore, when combined to {\tt
PDFFlow}~\cite{Carrazza:2020qwu,juan_cruz_martinez_2020_3964191}, it is possible
to perform a full simulation of physical processes with quite competitive
performance in comparison to specialized codes.

However, the integration of {\tt PineAPPL}, or any other external library that
expects input from {\tt VegasFlow}, may generate a natural performance
deterioration of the Monte Carlo simulation thanks to operations that may not
benefit from the multithreading paradigm of {\tt VegasFlow} on hardware
accelerators. {\tt PineAPPL} is designed for CPUs and provides to the developer
the possibility to be distributed in a multithreading configuration, however it
does not provide a GPU implementation and methods for asynchronous grid filling.

\begin{figure}
  \centering
  \includegraphics[width=0.5\textwidth]{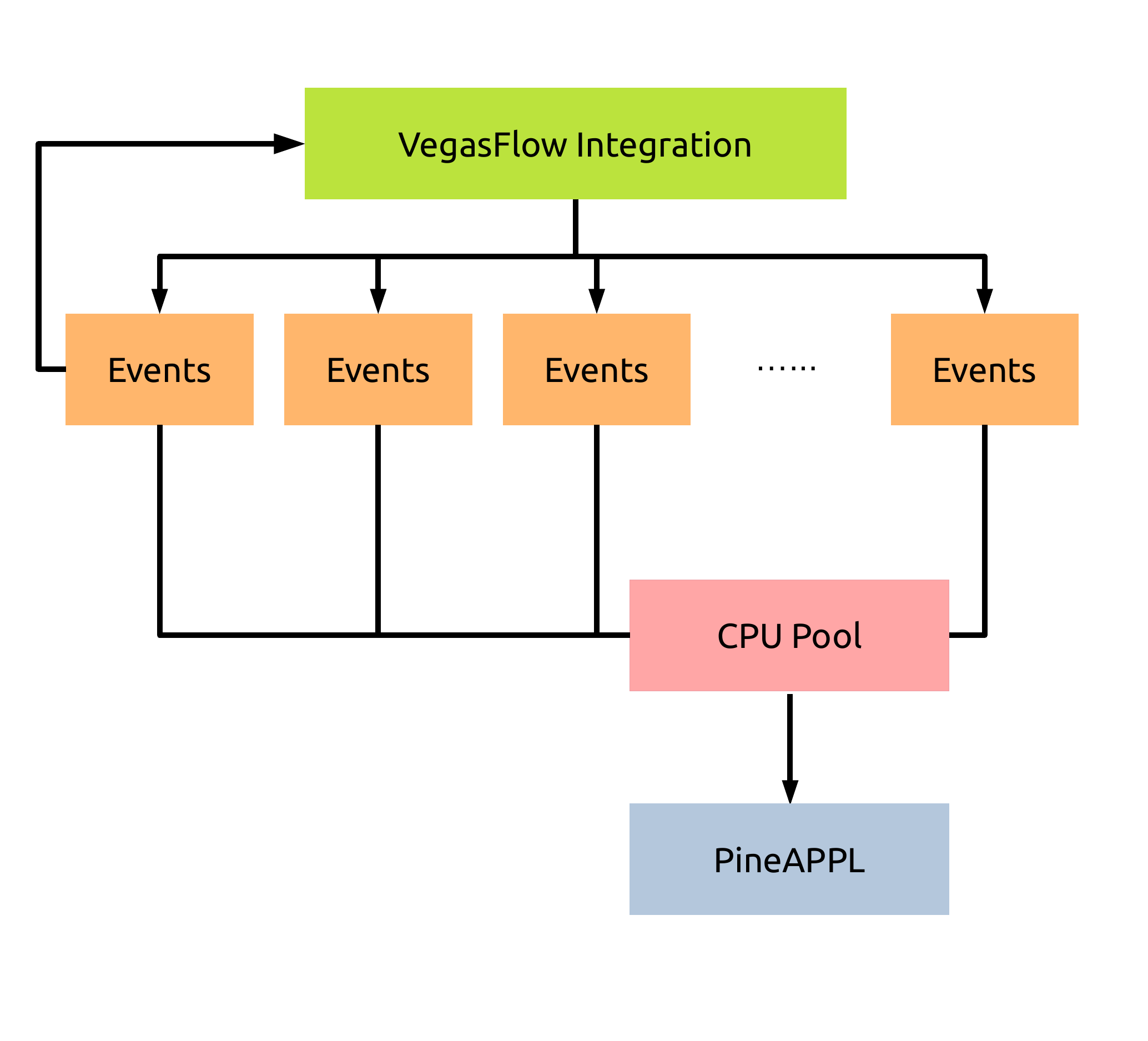}
  \caption{\label{fig:scheme}
  Schematic representation of the {\tt VegasFlow} and {\tt PineAPPL}
  integration. The evaluation of Monte Carlo events is driven by the {\tt
  VegasFlow} framework. At the end of each event, a separate CPU thread pool
  receives all the required information to fill a {\tt PineAPPL} grid. The
  procedure is asynchronous between the Monte Carlo integration and grid fill.}
\end{figure}

In order to provide a solution for such a problem, in Figure~\ref{fig:scheme} we
represent schematically the approach proposed here. The Monte Carlo simulation
is driven by the {\tt VegasFlow} framework, which takes care of generating
events and distributing them among available devices such as CPUs and/or GPUs.
Each batch of events evaluates a large number of matrix element weights and
phase space configurations which are subsequently sent to the {\tt PineAPPL}
library for the construction of fast-evaluation grids. In a multithreading CPU
and GPU environment, a sequential synchronous fill operation may reduce
drastically the performance of the Monte Carlo simulation, by increasing the
total amount of time needed to achieve precise predictions. In order to avoid
such a performance deterioration, we propose to detach the computation between the
{\tt VegasFlow} Monte Carlo event simulation from the {\tt PineAPPL} grid filling
operation by creating a CPU thread pool that asynchronously queues and executes
the operations required by {\tt PineAPPL}.

From a technical perspective all steps presented in the previous paragraph can
be achieved using the Python interfaces of {\tt VegasFlow} and {\tt PineAPPL}.
In particular, we can generate Monte Carlo events using the eager mode feature
in {\tt VegasFlow} and including an asynchronous job execution CPU thread pool
using the standard Python multiprocessing
module\footnote{\url{https://docs.python.org/3/library/multiprocessing.html\#module-multiprocessing}}.

\section{Measuring performance}

In order to test the asynchronous approach proposed in the previous section, we
use {\tt VegasFlow} with a simplified simulation of Drell--Yan, which only
calculates the leading-order matrix element for the photon-induced process,
$\gamma \gamma \to \ell \bar{\ell}$. The simple structure of the corresponding
matrix element and phase space allows to easily test a wide range of
phase-space points. Furthermore, the evaluation is cheap enough to highlight
the overhead of filling a {\tt PineAPPL} grid, which is basically a constant of
the number of partonic processes; when simulating more complex processes, we
therefore expect the relative overhead to be much lower.

At the end of each batch of events, we compute a {\tt PineAPPL} grid for the
$|y_{\ell\bar{\ell}}|$ observable. In terms of physical cuts we consider a
single invariant-mass slice of a CMS 8 TeV analysis~\cite{Chatrchyan:2013tia}, which
requires $60 < m_{\ell\bar{\ell}} < 120$ GeV, $p_{T}^{\ell} > 14$ GeV,
$|y_{\ell\bar{\ell}}| < 2.4$ and $|y_{\ell}| < 2.4$. From the {\tt
VegasFlow} point of view we tested the performance by increasing number of
events before cuts from $10^3$ to $10^9$, using the default maximum number of
events per device of $10^6$.

\begin{figure}
  \includegraphics[width=0.5\textwidth]{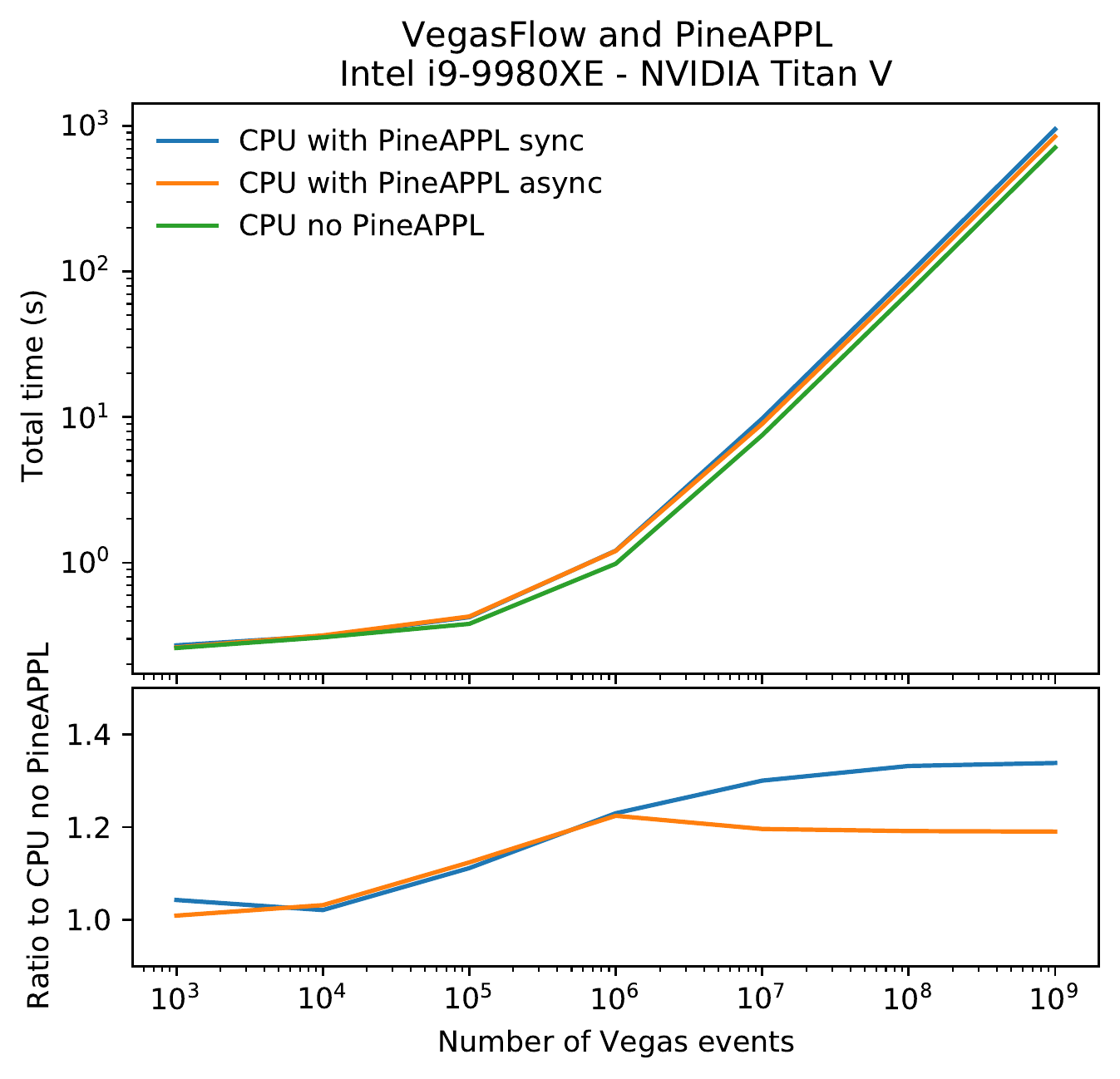}%
  \includegraphics[width=0.5\textwidth]{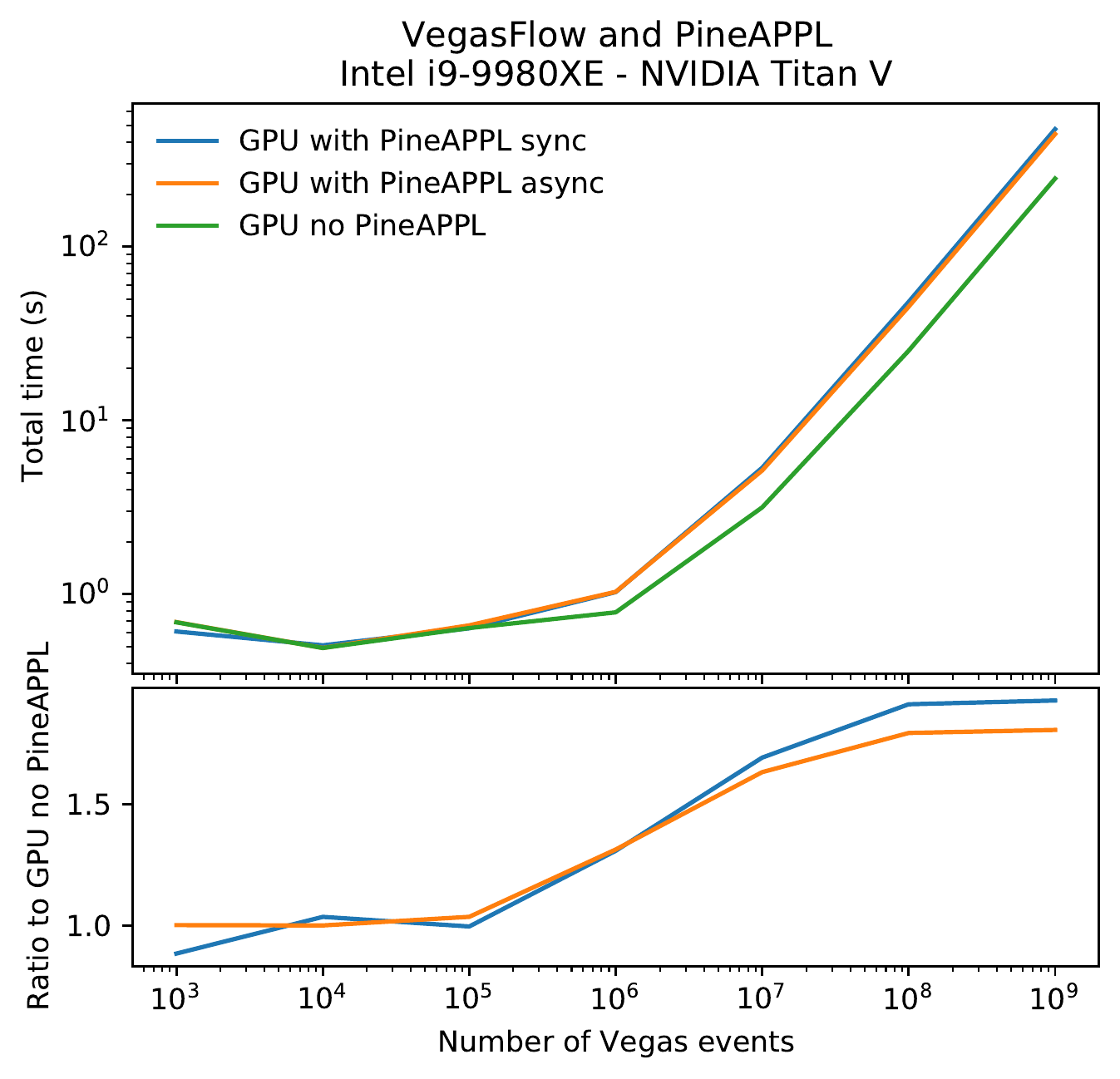}
  \caption{\label{fig:results}Performance results for the Monte Carlo simulation
  of photon-induced Drell--Yan channel using {\tt VegasFlow} and {\tt PineAPPL}
  on CPU (left) and GPU (right). The synchronous approach produces a performance
  deterioration up to $\approx 30$\% on CPU and $\approx 80$\% on GPU, while the
  asynchronous approach reduces the overall deterioration by $\approx 10$\%. Note
  that the asynchronous approach detaches the Monte Carlo integration from the
  grid filling operations providing further options for optimization.}
\end{figure}

In the left plot of Figure~\ref{fig:results} we show the total computing time of
the Monte Carlo simulation for an increasing number of events. We compare runs
without {\tt PineAPPL} (green) to the synchronous (blue) and asynchronous
(orange) approaches with {\tt PineAPPL}. Similarly, on the right plot of
Figure~\ref{fig:results}, we present performance results on GPU.
The synchronous approach produces a performance deterioration up to $\approx
30$\% on CPU and $\approx 80$\% on GPU, while the asynchronous approach reduces
the overall deterioration by $\approx 10$\%. The main advantage of the
asynchronous approach is the possibility to detach the Monte Carlo event
simulation from the operations related to the construction of {\tt PineAPPL}
grids. From the Monte Carlo simulation perspective we observe a negligible
overhead due to the submission of asynchronous jobs. On the other hand, the CPU
thread pool, used by the asynchronous job, can be further optimized
depending on the available system resources.
Note that the system used for the performance measurement in
Figure~\ref{fig:results} has a fast clock CPU thus the expect larger performance
deteriorations on average professional grade hardware.

\section{Outlook}

The example presented in this proceedings shows that an asynchronous approach to
filling fast-interpolation grids using {\tt PineAPPL} is feasible and provides a
useful interface to detach Monte Carlo event simulation from sequential
operations without introducing a strong performance deterioration.

We have compared the performance of synchronous and asynchronous approaches on
CPU and GPU for a leading-order Drell--Yan photon-induced simulation. As
expected, the performance from an asynchronous approach outperforms the
sequential synchronous mechanism and does not introduce overheads to the Monte
Carlo event simulation.

The code for the exercise presented in this manuscript is public available at
the {\tt VegasFlow}
repository~\cite{Carrazza:2020rdn,juan_cruz_martinez_2020_3691927} inside the
examples folder.

\section*{Acknowledgements}

We acknowledge the NVIDIA Corporation for the donation of a Titan V GPU used for
this research. This project is supported by the European Research Council under
the European Unions Horizon 2020 research and innovation Programme (grant
agreement number 740006) and by the UNIMI Linea2A project ``New hardware for
HEP''.


\newpage

\begin{thebibliography}{99}

\bibitem{Forte:2020yip}
S.~Forte and S.~Carrazza,
[arXiv:2008.12305 [hep-ph]].

\bibitem{Carli:2010rw}
T.~Carli, D.~Clements, A.~Cooper-Sarkar, C.~Gwenlan, G.~P. Salam, F.~Siegert
  et~al., \emph{{A posteriori inclusion of parton density functions in NLO QCD
  final-state calculations at hadron colliders: The APPLGRID Project}},
  \href{https://doi.org/10.1140/epjc/s10052-010-1255-0}{\emph{Eur.\ Phys.\ J.\
  C} {\bfseries 66} (2010) 503}
  [\href{https://arxiv.org/abs/0911.2985}{{\ttfamily 0911.2985}}].

\bibitem{Kluge:2006xs}
T.~Kluge, K.~Rabbertz and M.~Wobisch, \emph{{FastNLO: Fast pQCD calculations
for PDF fits}},  in \emph{{Deep inelastic scattering. Proceedings, 14th
International Workshop, DIS 2006, Tsukuba, Japan, April 20-24, 2006}},
pp.~483--486, 9, 2006,
\href{https://doi.org/10.1142/9789812706706\_0110}{DOI}
[\href{https://arxiv.org/abs/hep-ph/0609285}{{\ttfamily hep-ph/0609285}}].

\bibitem{Carrazza:2020gss}
S.~Carrazza, E.~R.~Nocera, C.~Schwan and M.~Zaro,
[arXiv:2008.12789 [hep-ph]].

\bibitem{christopher_schwan_2020_3992765}
C.~Schwan and S.~Carrazza,
\href{https://doi.org/10.5281/zenodo.3992765}{N3PDF/PineAPPL} (Aug. 2020).
\newblock \href {https://doi.org/10.5281/zenodo.3992765}
  {\path{doi:10.5281/zenodo.3992765}}.

\bibitem{Carrazza:2020rdn}
S.~Carrazza and J.~M.~Cruz-Martinez,
Comput. Phys. Commun. \textbf{254} (2020), 107376
doi:10.1016/j.cpc.2020.107376
[arXiv:2002.12921 [physics.comp-ph]].

\bibitem{juan_cruz_martinez_2020_3691927}
J.~Cruz-Martinez, S.~Carrazza,
\href{https://doi.org/10.5281/zenodo.3691926}{N3PDF/VegasFlow} (Feb. 2020).
\newblock \href {https://doi.org/10.5281/zenodo.3691926}
  {\path{doi:10.5281/zenodo.3691926}}.

\bibitem{Carrazza:2020qwu}
S.~Carrazza, J.~M.~Cruz-Martinez and M.~Rossi,
[arXiv:2009.06635 [hep-ph]].

\bibitem{juan_cruz_martinez_2020_3964191}
J.~Cruz-Martinez, M.~Rossi, S.~Carrazza,
  \href{https://doi.org/10.5281/zenodo.3964191}{N3PDF/PDFFlow} (Jul. 2020).
\newblock \href {https://doi.org/10.5281/zenodo.3964191}
  {\path{doi:10.5281/zenodo.3964191}}.

\bibitem{Chatrchyan:2013tia}
S.~Chatrchyan \textit{et al.} [CMS],
JHEP \textbf{12} (2013), 030
doi:10.1007/JHEP12(2013)030
[arXiv:1310.7291 [hep-ex]].

\end{thebibliography}
\end{document}